\documentclass[aps,prd,onecolumn,nofootinbib]{revtex4-2}

\usepackage{amsmath,amssymb}
\usepackage{tikz}
\usepackage{xcolor}
\usepackage{hyperref}
\hypersetup{hidelinks}
\begin{document}

\title{Holographic Dual of Synge's Identity in AdS/CFT}

\author{Bingbing Chen}
\email{chenbingbing@scun.edu.cn}
\affiliation{School of Mathematics, Physics and Statistics, Sichuan Minzu College, Kangding 626001,  China}

\author{Deyou Chen}
\email{deyouchen@xhu.edu.cn}
\affiliation{School of Science, Xihua University, Chengdu 610039,  China}


\begin{abstract}
We identify CFT counterparts of Synge's identity for finite AdS geodesics in arbitrary dimensions. Starting from exact finite bulk--boundary pairs, the  Synge equation induces endpoint scale equations for a Weyl-frame scalar two-point function and for the finite entanglement entropy between disjoint balls. In the planar frame, the common-scale correlator equation is the dilatation Ward identity. In the entanglement sector, the corresponding relative-scale equation is governed by a transverse hyperbolic area and reduces in two dimensions to the previously identified entanglement RG equation. 
\end{abstract}

\maketitle

\section{Introduction}

Holographic duality is usually formulated as a dictionary between boundary observables and bulk quantities \cite{Maldacena:1997re,Gubser:1998bc,Witten:1998qj}. Correspondences between equations on the two sides are also familiar; for example, bulk constraints generate boundary Ward identities in Hamilton-Jacobi formulations of holographic renormalization \cite{Corley:2000ct,Martelli:2002sp}. The question considered here is more specific. Suppose that a finite bulk observable and a CFT observable are already related exactly. Can a differential identity obeyed by the geometric member be transported directly to a CFT equation? A distinctive feature of this setting is that the observable pair itself is {\it finite and exact}, so this translation does not require an additional boundary regulator, a heavy-operator limit, or a geodesic-saddle approximation. The resulting simplicity is therefore a consequence of the starting point rather than of a further approximation.

The geometric identity of interest in this work is Synge's identity. For two points connected by a geodesic of length $\ell$, Synge's world function is \cite{Synge:1960ueh,Poisson:2011nh}
\begin{equation}
\sigma=\frac{1}{2}\ell^2,
\label{eq:sigma}
\end{equation}
and obeys
\begin{equation}
g^{\mu\nu}\partial_{\mu}\sigma\,\partial_{\nu}\sigma=2\sigma,
\label{eq:synge}
\end{equation}
where the derivatives act on either  of the end points.
Away from coincidence this is equivalently
\begin{equation}
g^{\mu\nu}\partial_{\mu}\ell\,\partial_{\nu}\ell=1,
\label{eq:hj}
\end{equation}
the Hamilton--Jacobi, or eikonal, equation for unit-speed geodesic flow \cite{Katanaev:2023geodesic}. In AdS$_3$/CFT$_2$, Ref.~\cite{Jiang:2024hjz} showed that it is conformally equivalent to the renormalization-group equation obeyed by the relevant finite disjoint entanglement entropy. The higher-dimensional question remained open because the CFT observable paired with a generic finite bulk geodesic had not yet been identified.

The required finite observables are now available. Ref.~\cite{Jiang:2025jnk} obtained the finite disjoint entanglement entropy between disjoint balls in arbitrary-dimensional CFT and its exact relation to the corresponding entanglement wedge cross-section (EWCS). Ref.~\cite{Jiang:2026juf} established an exact pair between a scalar two-point function in the Weyl frame of an open solid torus and a finite geodesic lying entirely in the associated AdS interior. These relations were subsequently organized as an exact holographic kinematic sector in Ref.~\cite{Yang:2026Kinematics}. 

In this paper, we use these finite pairs to identify the CFT equations induced by Synge's identity. The derivation is performed first in a symmetric open-solid-torus representative, where the relevant bulk geodesic is radial and the relation between CFT and bulk endpoints is explicit. In the scalar-correlator sector, Synge's identity produces two endpoint equations whose common-scale combination becomes the planar dilatation Ward identity, while their difference retains the finite relative-scale information. A more interesting application concerns the finite disjoint entanglement entropy  $S_{\rm disj}$. The finite endpoint and relative-scale equations for $S_{\rm disj}$ are yielded. Both the correlator and entanglement equations can then be expressed entirely in terms of the conformally invariant inversive product $\varrho$, giving compact first-order equations for general disjoint spherical configurations. In two dimensions the entanglement equation reduces to the RG equation of Ref.~\cite{Jiang:2024hjz}; in higher dimensions its adjacent limit reproduces the familiar area-law scaling, consistent with the Ryu--Takayanagi relation \cite{Ryu:2006bv,Ryu:2006ef}.

\section{Open solid torus and the exact finite pair}
\label{sec:pair}

\begin{figure}[h]
\centering
\includegraphics[scale=0.5]{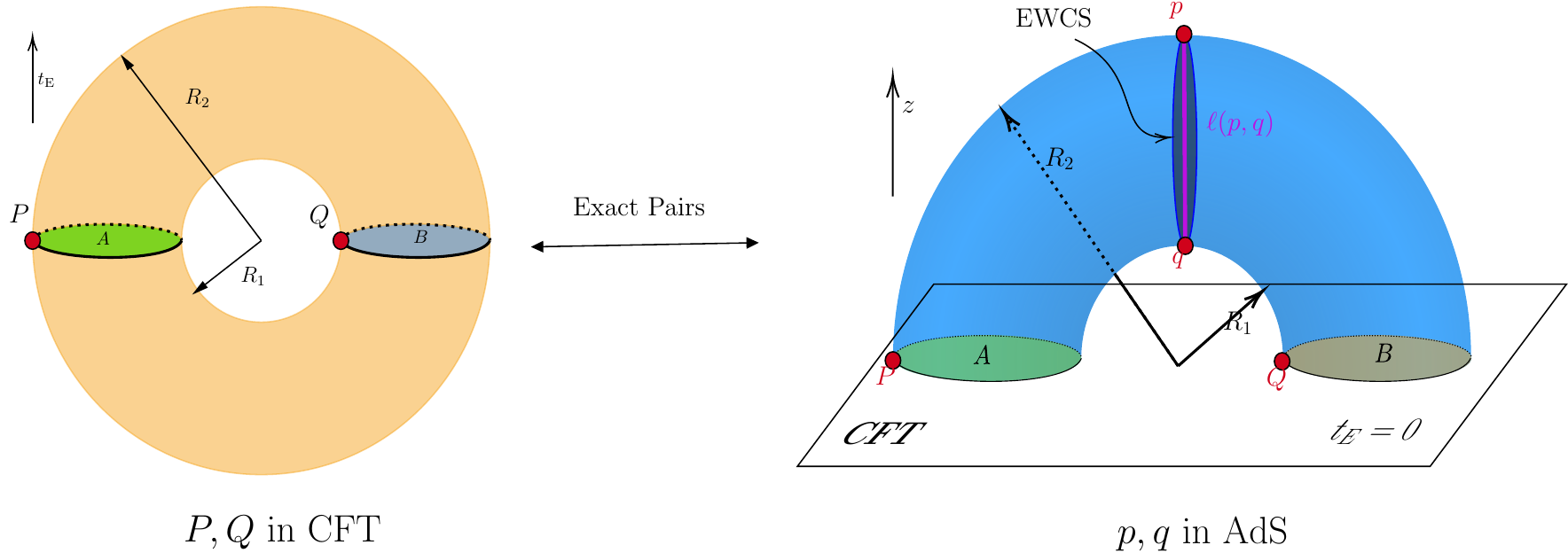}
\caption{Schematic symmetric exact pair. The CFT/open-solid-torus geometry (left) and the EAdS bulk geometry (right) are  shown with different colors. The CFT points $P,Q$ and the bulk points $p,q$ are distinct. In the symmetric representative the dictionary identifies $z_p=R_1$ and $z_q=R_2$, giving $L(P,Q)=\ell(p,q)=\log(R_2/R_1)$.}
\label{fig:pair}
\end{figure}

We work in Euclidean signature and set the AdS radius to unity. The CFT is placed on the flat open solid torus \cite{Jiang:2025jnk,Jiang:2026juf}
\begin{equation}
\mathcal{B}_{D}
=
\left\{
\left(\sqrt{t_E^2+y^2}-\frac{R_2+R_1}{2}\right)^2
+|\boldsymbol{x}|^2
<
\left(\frac{R_2-R_1}{2}\right)^2
\right\},
\qquad R_2>R_1>0,
\label{eq:ost}
\end{equation}
as illustrated in Fig. \ref{fig:pair}, left panel.
With $t_E=r\sin\theta$ and $y=r\cos\theta$, the Weyl-rescaled metric is
\begin{equation}
ds_{\mathcal W}^2
=
d\theta^2+\frac{dr^2+d\boldsymbol{x}^{\,2}}{r^2}.
\label{eq:weylmetric}
\end{equation}
We choose the CFT insertion points
\begin{equation}
P=(R_2,\pi,\boldsymbol{0}),
\qquad
Q=(R_1,0,\boldsymbol{0}),
\label{eq:PQ}
\end{equation}
where $P,Q$ will always denote CFT points. Their Weyl-frame hyperbolic separation is
\begin{equation}
L(P,Q)=\log\frac{R_2}{R_1}.
\label{eq:L}
\end{equation}
The associated Euclidean AdS$_{D+1}$ metric is
\begin{equation}
ds_{\rm EAdS}^2
=
\frac{dz^2+dt_E^2+d\boldsymbol{x}^{\,2}}{z^2}.
\label{eq:adsmetric}
\end{equation}
Let $p,q$ denote the antipodal bulk points on the entanglement wedge cross-section (EWCS) that define the finite geodesic pair, as illustrated by the right panel of Fig. \ref{fig:pair}. In the symmetric representative their radial positions satisfy \cite{Jiang:2026juf}
\begin{equation}
z_p=R_1,
\qquad
z_q=R_2,
\label{eq:zmap}
\end{equation}
and hence
\begin{equation}
\ell(p,q)
=
\int_{z_p}^{z_q}\frac{dz}{z}
=
\log\frac{R_2}{R_1}
=
L(P,Q).
\label{eq:exactdistance}
\end{equation}
Equation~\eqref{eq:exactdistance} is an observable-level dictionary statement. It does not identify the CFT points $P,Q$ with the bulk points $p,q$, nor does it identify derivatives taken in the two spaces.

It is useful to separate the common and relative scales,
\begin{equation}
R=\sqrt{R_1R_2},
\qquad
\eta=\frac{1}{2}\log\frac{R_2}{R_1}=\frac{\ell}{2}.
\label{eq:scales}
\end{equation}
A common dilation changes $R$ while leaving $\eta$ invariant. Thus $\ell$ is represented on the CFT side by a relative logarithmic scale rather than by the overall RG scale.

For two general spheres of radii $r,r'$ and centers $\boldsymbol{x},\boldsymbol{x}'$, the conformal class is characterized by the inversive product \cite{beardon2012geometry}
\begin{equation}
\varrho
=
\left|
\frac{r^2+r'^2-|\boldsymbol{x}-\boldsymbol{x}'|^2}{2rr'}
\right|.
\label{eq:varrho}
\end{equation}
The symmetric representative gives
\begin{equation}
\ell(\varrho)
=
\log
\frac{\sqrt{\varrho+1}+\sqrt{2}}
{\sqrt{\varrho+1}-\sqrt{2}},
\label{eq:ellrho}
\end{equation}
with
\begin{equation}
\tanh^2\eta=\frac{2}{\varrho+1},
\qquad
\sinh^2\eta=\frac{2}{\varrho-1}.
\label{eq:etarho}
\end{equation}
We will first derive the CFT equations in the symmetric radial representative and only afterwards rewrite those equations in terms of $\varrho$.

\section{Synge's identity in the radial representative}
\label{sec:synge_radial}

Synge's identity is applied only to the genuine bulk geodesic. With the ordering $z_q>z_p$, the radial Hamilton--Jacobi equation fixes the physical branches
\begin{equation}
z_p\frac{\partial\ell}{\partial z_p}=-1,
\qquad
z_q\frac{\partial\ell}{\partial z_q}=+1.
\label{eq:bulkbranches}
\end{equation}
The opposite ordering simply exchanges the endpoints. Restricting to the exact-pair family $z_p=R_1$, $z_q=R_2$ gives
\begin{equation}
R_1\frac{\partial\ell}{\partial R_1}=-1,
\qquad
R_2\frac{\partial\ell}{\partial R_2}=+1.
\label{eq:cftbranches}
\end{equation}
The derivatives in Eq.~\eqref{eq:cftbranches} act on the CFT/open-solid-torus parameters. Equation~\eqref{eq:zmap} is what pulls the bulk endpoint relation back to this symmetric CFT family.

\section{Two-point-function counterpart}
\label{sec:twopt}

\subsection{Weyl-frame endpoint equations}

For a scalar primary of dimension $\Delta$, the exact finite pair is \cite{Jiang:2026juf}
\begin{equation}
G_{\mathcal W}(P,Q)
=
\frac{C_{\Delta}}
{\left[2\cosh(\ell/2)\right]^{2\Delta}}.
\label{eq:exactpair}
\end{equation}
Combining Eq.~\eqref{eq:exactpair} with the Synge branches \eqref{eq:cftbranches} gives the CFT endpoint equations
\begin{equation}
R_1\frac{\partial}{\partial R_1}\log G_{\mathcal W}
=
\Delta\tanh\frac{\ell}{2},
\label{eq:gw1}
\end{equation}
and
\begin{equation}
R_2\frac{\partial}{\partial R_2}\log G_{\mathcal W}
=
-\Delta\tanh\frac{\ell}{2}.
\label{eq:gw2}
\end{equation}
Their sum gives common-scale invariance,
\begin{equation}
\left(
R_1\frac{\partial}{\partial R_1}
+
R_2\frac{\partial}{\partial R_2}
\right)G_{\mathcal W}=0,
\label{eq:gwcommon}
\end{equation}
while their difference gives, with $\eta=\ell/2$,
\begin{equation}
\frac{\partial}{\partial\eta}\log G_{\mathcal W}
=
-2\Delta\tanh\eta.
\label{eq:gweta}
\end{equation}
Equations~\eqref{eq:gw1}--\eqref{eq:gweta}, rather than the closed-form correlator itself, are the direct Weyl-frame CFT image of the radial Synge identity.

\subsection{Planar frame and the dilatation Ward identity}

In the original planar frame the same CFT insertion points are located at $P=(t_E=0,y=-R_2,\boldsymbol{0})$ and $Q=(t_E=0,y=R_1,\boldsymbol{0})$. Weyl covariance gives
\begin{equation}
G_{\mathcal W}(P,Q)
=
(R_1R_2)^{\Delta}G_{\rm pl}(P,Q).
\label{eq:weylplanar}
\end{equation}
Equations~\eqref{eq:gw1} and \eqref{eq:gw2} therefore become
\begin{equation}
R_1\frac{\partial}{\partial R_1}\log G_{\rm pl}
=
-\Delta+\Delta\tanh\frac{\ell}{2},
\label{eq:gpl1}
\end{equation}
and
\begin{equation}
R_2\frac{\partial}{\partial R_2}\log G_{\rm pl}
=
-\Delta-\Delta\tanh\frac{\ell}{2}.
\label{eq:gpl2}
\end{equation}
Their sum is
\begin{equation}
\left(
R_1\frac{\partial}{\partial R_1}
+
R_2\frac{\partial}{\partial R_2}
+2\Delta
\right)G_{\rm pl}=0,
\label{eq:wardradial}
\end{equation}
which is precisely the restriction to the symmetric representative of the standard planar dilatation Ward identity \cite{Osborn:1993cr},
\begin{equation}
\left(
x_1^{\mu}\frac{\partial}{\partial x_1^{\mu}}
+
x_2^{\mu}\frac{\partial}{\partial x_2^{\mu}}
+2\Delta
\right)
G_{\rm pl}(x_1,x_2)=0.
\label{eq:wardfull}
\end{equation}
The difference of Eqs.~\eqref{eq:gpl1} and \eqref{eq:gpl2} instead retains the finite relative-scale information,
\begin{equation}
\left(
R_2\frac{\partial}{\partial R_2}
-
R_1\frac{\partial}{\partial R_1}
\right)
\log G_{\rm pl}
=
-2\Delta\tanh\frac{\ell}{2}.
\label{eq:gplrelative}
\end{equation}
These equations are not additional CFT dynamics. Their significance is the identification of the precise CFT relations induced by Synge's identity under the finite exact pair.

\section{Entanglement-entropy counterpart}
\label{sec:entropy}

\subsection{Exact entropy relation}

Referring to the left panel of Fig. \ref{fig:pair}, the entanglement entropy $S_{\rm disj}(A:B)$ between  disjoint spacelike complementary regions $A$ and $B$ are calculated in ref.~\cite{Jiang:2025jnk}, 
\begin{equation}
S_{\rm disj}(A:B)
=
-4\pi\mathcal{E}_{\rm vac}\,
\operatorname{Vol}(\mathbb{B}^{D-1}),
\label{eq:Sknown}
\end{equation}
where $\mathcal{E}_{\rm vac}<0$ is the Weyl-frame vacuum energy density. This entanglement entropy is exactly paired with the EWCS volume \cite{Jiang:2025jnk,Yang:2026Kinematics}. In a semiclassical holographic realization the normalization it reduces to the usual RT normalization \cite{Ryu:2006bv,Ryu:2006ef}. 
This finite disjoint entanglement entropy can equivalently be written in the symmetric representative as
\begin{equation}
S_{\rm disj}(\ell)
=
4\pi\mathcal{E}_{D}\,\Omega_{D-2}
\int_0^{\ell/2}\sinh^{D-2}u\,du,
\label{eq:Sell}
\end{equation}
where
\begin{equation}
\Omega_{D-2}
=
\frac{2\pi^{(D-1)/2}}
{\Gamma\left((D-1)/2\right)}\qquad {\rm and}\qquad \mathcal{E}_{D}=|\mathcal{E}_{\rm vac}|,
\label{eq:Omega}
\end{equation}
The exact   relation \eqref{eq:Sell} has the differential form
\begin{equation}
\frac{dS_{\rm disj}}{d\ell}
=
2\pi\mathcal{E}_{D}\,\Omega_{D-2}
\sinh^{D-2}\frac{\ell}{2}.
\label{eq:Sflow_observable}
\end{equation}
This equation follows from the known entropy as a function of $\ell$; by itself it is not yet the CFT image of Synge's identity.

\subsection{Synge-induced endpoint and relative-scale equations}

Combining the observable relation \eqref{eq:Sflow_observable} with the Synge branches \eqref{eq:cftbranches} gives
\begin{equation}
R_1\frac{\partial S_{\rm disj}}{\partial R_1}
=
-2\pi\mathcal{E}_{D}\,\Omega_{D-2}
\sinh^{D-2}\frac{\ell}{2},
\label{eq:S1}
\end{equation}
and
\begin{equation}
R_2\frac{\partial S_{\rm disj}}{\partial R_2}
=
+2\pi\mathcal{E}_{D}\,\Omega_{D-2}
\sinh^{D-2}\frac{\ell}{2}.
\label{eq:S2}
\end{equation}
These endpoint equations are the direct entanglement counterpart of Synge's identity. Their sum gives
\begin{equation}
\left(
R_1\frac{\partial}{\partial R_1}
+
R_2\frac{\partial}{\partial R_2}
\right)S_{\rm disj}=0,
\label{eq:Scommon}
\end{equation}
while their difference gives
\begin{equation}
\left(
R_2\frac{\partial}{\partial R_2}
-
R_1\frac{\partial}{\partial R_1}
\right)S_{\rm disj}
=
4\pi\mathcal{E}_{D}\,\Omega_{D-2}
\sinh^{D-2}\frac{\ell}{2}.
\label{eq:Srelative}
\end{equation}
Equivalently,
\begin{equation}
\frac{\partial S_{\rm disj}}{\partial\eta}
=
4\pi\mathcal{E}_{D}\,\Omega_{D-2}
\sinh^{D-2}\eta.
\label{eq:Seta}
\end{equation}
The common-scale equation expresses conformal scale invariance, whereas Eq.~\eqref{eq:Seta} is the Synge-induced finite relative-scale evolution equation.

\section{Conformally invariant form for general configurations}
\label{sec:general}

The equations above were derived in the symmetric radial representative. We now convert the Synge-induced relative-scale equations themselves to the conformal invariant $\varrho$. From Eq.~\eqref{eq:etarho}, 
\begin{equation}
\frac{d\varrho}{d\eta}
=
-(\varrho^2-1)\tanh\eta.
\label{eq:drhodeta}
\end{equation}
This is only a change from the relative-scale coordinate $\eta$ to the invariant labeling the full conformal class.

For the two-point function, combining Eqs.~\eqref{eq:gweta} and \eqref{eq:drhodeta} gives
\begin{equation}
(\varrho^2-1)
\frac{d}{d\varrho}\log G_{\mathcal W}(\varrho)
=
2\Delta.
\label{eq:GODE}
\end{equation}
Thus Eq.~\eqref{eq:GODE} is the conformally invariant form of the Synge-induced relative-scale equation, rather than a derivative taken from an independently inserted closed-form correlator. Integrating it recovers the exact finite result
\begin{equation}
G_{\mathcal W}(\varrho)
=
\frac{C_{\Delta}}{4^{\Delta}}
\left(
\frac{\varrho-1}{\varrho+1}
\right)^{\Delta}.
\label{eq:Gwrho_solution}
\end{equation}

For the entanglement entropy, Eqs.~\eqref{eq:Seta}, \eqref{eq:drhodeta}, and \eqref{eq:etarho} give
\begin{equation}
\frac{dS_{\rm disj}}{d\varrho}
=
-2\pi\mathcal{E}_{D}\,\Omega_{D-2}
\frac{2^{(D-1)/2}}
{(\varrho-1)^{D/2}\sqrt{\varrho+1}}.
\label{eq:SODE}
\end{equation}
This is the invariant form of the Synge-induced entanglement equation. It no longer refers to the symmetric representative and therefore applies to general disjoint spherical configurations in the same conformal class. Integrating Eq.~\eqref{eq:SODE} with the physical large-separation condition
\begin{equation}
S_{\rm disj}(\varrho\rightarrow\infty)=0
\label{eq:Sboundary}
\end{equation}
reproduces the hypergeometric entropy of Ref.~\cite{Jiang:2025jnk}. In this sense the previously known exact observables are recovered as integrated solutions of the invariant equations generated from the radial Synge relation.

\section{Two dimensions and the cavity limit}
\label{sec:limits}

For $D=2$,
\begin{equation}
\Omega_0=2,
\qquad
\mathcal{E}_2=\frac{c}{24\pi},
\label{eq:D2data}
\end{equation}
so the exact entropy relation becomes
\begin{equation}
\frac{dS_{\rm disj}}{d\ell}=\frac{c}{6},
\label{eq:D2ell}
\end{equation}
and the Synge-induced endpoint equations reduce to
\begin{equation}
R_1\frac{\partial S_{\rm disj}}{\partial R_1}=-\frac{c}{6},
\qquad
R_2\frac{\partial S_{\rm disj}}{\partial R_2}=+\frac{c}{6}.
\label{eq:D2endpoints}
\end{equation}
There is a minor but important notational difference from Ref.~\cite{Jiang:2024hjz}. There the symbol $\ell$ denotes the RG length scale itself. To avoid confusion, denote that scale here by $\lambda$. Holding the other endpoint fixed at a reference scale $\lambda_0$, the present geodesic length is
\begin{equation}
\ell_{\rm geo}=\log\frac{\lambda}{\lambda_0},
\label{eq:logRG}
\end{equation}
so that
\begin{equation}
\frac{d}{d\ell_{\rm geo}}
=
\lambda\frac{d}{d\lambda}.
\label{eq:RGderivative}
\end{equation}
In particular, choosing $\lambda=R_2$ and keeping $R_1=\lambda_0$ fixed, the second Synge-induced endpoint equation in Eq.~\eqref{eq:D2endpoints} becomes $\lambda\,dS/d\lambda=c/6$, precisely the RG equation of Ref.~\cite{Jiang:2024hjz}. Equivalently it may be written as Eq.~\eqref{eq:D2ell} in terms of the geodesic length. The two-dimensional result is the special case in which the transverse sphere has dimension zero and hence constant area. For $D>2$ the transverse area grows as $\sinh^{D-2}(\ell/2)$, so the corresponding statement is naturally a finite relative-scale evolution equation rather than a linear RG equation.

\begin{figure}[h]
\centering
\includegraphics[scale=1]{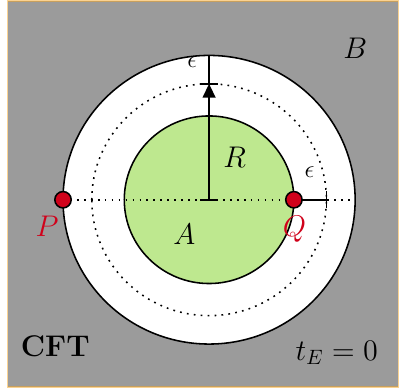}
\caption{Schematic cavity degeneration. As $\epsilon\to 0$, the disjoint configuration reduces to the usual adjacent configuration. The  finite scale equations develop the standard logarithmic ($d=2$) or area-law behavior ($d>2$).}
\label{fig:cavity}
\end{figure}

The cavity configuration provides the universal singular limit from the finite open-solid-torus description to the usual planar adjacent setup \cite{Jiang:2025jnk,Jiang:2026juf,Yang:2026Kinematics}. Referring to Fig. \ref{fig:cavity}, for concentric spherical scales
\begin{equation}
r=R+\epsilon,
\qquad
r'=R-\epsilon,
\label{eq:cavityradii}
\end{equation}
one has
\begin{equation}
\varrho
=
\frac{R^2+\epsilon^2}{R^2-\epsilon^2}.
\label{eq:cavityrho}
\end{equation}
Therefore
\begin{equation}
\sinh\frac{\ell}{2}
=
\frac{\sqrt{R^2-\epsilon^2}}{\epsilon}
\sim
\frac{R}{\epsilon},
\qquad \epsilon\rightarrow0.
\label{eq:cavityell}
\end{equation}
The Synge-induced endpoint equations then inherit the same short-distance scaling as the exact entropy relation,
\begin{equation}
\left|
R_i\frac{\partial S_{\rm disj}}{\partial R_i}
\right|
\sim
2\pi\mathcal{E}_{D}\,\Omega_{D-2}
\left(\frac{R}{\epsilon}\right)^{D-2}.
\label{eq:arealimit}
\end{equation}
For $D=2$ this integrates to the logarithmic law; for $D>2$ it displays the standard area-law scaling. This agrees with the adjacent limit obtained in Ref.~\cite{Jiang:2025jnk} and, in the semiclassical holographic regime, with the RT result \cite{Ryu:2006bv,Ryu:2006ef}.

\section{Discussion and conclusion}
\label{sec:discussion}

The logical structure of the result is simple. The exact finite pairs first supply observable relations such as $G_{\mathcal W}=G_{\mathcal W}(\ell)$ and $S_{\rm disj}=S_{\rm disj}(\ell)$. Synge's identity then supplies the endpoint derivatives of the genuine bulk geodesic. Combining the two produces the CFT endpoint equations \eqref{eq:gw1}, \eqref{eq:gw2}, \eqref{eq:S1}, and \eqref{eq:S2}, together with their common- and relative-scale combinations. In the planar correlator sector the common-scale equation is the dilatation Ward identity. In the entanglement sector the relative-scale equation is controlled by a transverse hyperbolic area. Finally, changing variables from $\eta$ to conformal invariant
$\varrho$ converts these Synge-induced equations into the invariant equations \eqref{eq:GODE} and \eqref{eq:SODE} for general disjoint spherical configurations.

The distinction between these steps matters. Equations such as \eqref{eq:Sflow_observable} are differential forms of already known exact observable relations; they do not by themselves use Synge's identity. The genuine CFT counterparts of Synge's identity are the endpoint and common/relative-scale equations obtained after the bulk Hamilton--Jacobi branches are combined with those observable relations. The invariant $\varrho$ equations are then their conformally invariant continuation. Written in this order, the previously known closed forms as functions of $\varrho$ are recovered as integrated solutions and consistency checks, rather than differentiated to manufacture the invariant equations.

The derivations are deliberately short because the underlying observable pairs are already finite and exact. No additional boundary regulator, heavy-operator limit, or geodesic saddle is introduced when the bulk identity is transported to CFT variables. This illustrates a practical use of exact finite pairs: once an {\it exact} observable-level map is established, relations obeyed by the paired observables can be studied directly rather than reconstructed through an asymptotic limiting prescription.

\end{document}